\documentclass[12pt]{article}
\usepackage{sw20jart}
\usepackage{amsmath}

\newtheorem{theorem}{Theorem}[section]

\newtheorem{lemma}[theorem]{Lemma}

\newtheorem{remark}[theorem]{Remark}

\newenvironment{proof}[1][Proof]{\noindent\textbf{#1.} }{\ \rule{0.5em}{0.5em}}
\allowdisplaybreaks
\input{tcilatex}

\begin{document}

\author{Mark Korenblit \\
Holon Institute of Technology, Israel\\
korenblit@hit.ac.il\bigskip \\
Vadim E. Levit\\
Ariel University, Israel\\
levitv@ariel.ac.il }
\title{A Solution of Simultaneous Recurrences}
\date{}
\maketitle

\section{Introduction}

There exist problems whose solving requires a solution of simultaneous
recurrences, such as, for example, the following ones:%
\begin{equation}
\left\{ 
\begin{array}{l}
a_{x}=2a_{x-1}+4b_{x-1}+1 \\ 
b_{x}=a_{x-1}+3b_{x-1}+2c_{x-1}+1 \\ 
c_{x}=2b_{x-1}+4c_{x-1}+1\text{ ,}%
\end{array}%
\right.  \label{recf32}
\end{equation}%
\begin{equation}
\left\{ 
\begin{array}{l}
a_{x}=\frac{37}{6}a_{x-1}-\frac{1}{6}b_{x-1}+2 \\ 
\\ 
b_{x}=\frac{15}{2}a_{x-1}-\frac{7}{2}b_{x-1}+2c_{x-1}+2 \\ 
\\ 
c_{x}=\frac{16}{3}a_{x-1}-\frac{10}{3}b_{x-1}+4c_{x-1}+2%
\end{array}%
\right.  \label{recf33}
\end{equation}%
Besides, the problem of solving simultaneous recurrences is of interest
itself, regardless of any particular application. Section \ref%
{sec_gen_sim_rec} discusses transformation of matrix recurrences to regular
recurrences. Section \ref{sec_2_3_sim_rec} describes a process of solving
special matrix recurrences of order three (as (\ref{recf32}) and (\ref%
{recf33})) by their decomposition to matrix recurrences of order two.

\section{Transformation of Matrix Recurrences to Regular Recurrences\label%
{sec_gen_sim_rec}}

\textit{Matrix recurrences} look as 
\begin{equation}
\left\{ 
\begin{array}{l}
y_{1_{x}}=\alpha _{11}y_{1_{x-1}}+\alpha _{12}y_{2_{x-1}}+\alpha
_{13}y_{3_{x-1}}+\ldots +\alpha _{1n}y_{n_{x-1}}+\alpha _{1} \\ 
y_{2_{x}}=\alpha _{21}y_{1_{x-1}}+\alpha _{22}y_{2_{x-1}}+\alpha
_{23}y_{3_{x-1}}+\ldots +\alpha _{2n}y_{n_{x-1}}+\alpha _{2} \\ 
y_{3_{x}}=\alpha _{31}y_{1_{x-1}}+\alpha _{32}y_{2_{x-1}}+\alpha
_{33}y_{3_{x-1}}+\ldots +\alpha _{3n}y_{n_{x-1}}+\alpha _{3} \\ 
\cdot \text{\qquad }\cdot \text{\qquad }\cdot \text{\qquad }\cdot \text{%
\qquad }\cdot \text{\qquad }\cdot \text{\qquad }\cdot \text{\qquad }\cdot 
\text{\qquad }\cdot \text{\qquad }\cdot \\ 
y_{n_{x}}=\alpha _{n1}y_{1_{x-1}}+\alpha _{n2}y_{2_{x-1}}+\alpha
_{n3}y_{3_{x-1}}+\ldots +\alpha _{nn}y_{n_{x-1}}+\alpha _{n}.%
\end{array}%
\right.  \label{recf14}
\end{equation}%
Here $x$ is a current number of variables $y_{1},y_{2},\ldots ,y_{n}$ ($%
\alpha $ followed by an index denotes a constant coefficient) in their
sequences generated by means of (\ref{recf14}) so that $y_{1_{0}},y_{2_{0}},%
\ldots ,y_{n_{0}}$ are initial values of $y_{1},y_{2},\ldots ,y_{n}$,
respectively. We intend to separate variables of these recurrences so that
each variable $y_{i}$ is expressed in previous values of $y_{i}$ only, i.e. 
\begin{equation}
y_{i_{x}}=\beta _{1}y_{i_{x-1}}+\beta _{2}y_{i_{x-2}}+\beta
_{3}y_{i_{x-3}}+\ldots  \label{recf34}
\end{equation}%
where $\beta _{j}$ $(j=1,2,3,\ldots )$ is a constant coefficient. Recurrence
(\ref{recf34}) is a \textit{regular recurrence}. Methods for solving regular
recurrences, providing explicit formulae, are presented in \cite{Ros}.

First, we solve the problem for the special case when there are no absolute
terms ($\alpha _{1}=\alpha _{2}=\ldots =\alpha _{n}=0$). In this case (\ref%
{recf14}) is reduced to 
\begin{equation}
\left\{ 
\begin{array}{l}
y_{1_{x}}=\alpha _{11}y_{1_{x-1}}+\alpha _{12}y_{2_{x-1}}+\alpha
_{13}y_{3_{x-1}}+\ldots +\alpha _{1n}y_{n_{x-1}} \\ 
y_{2_{x}}=\alpha _{21}y_{1_{x-1}}+\alpha _{22}y_{2_{x-1}}+\alpha
_{23}y_{3_{x-1}}+\ldots +\alpha _{2n}y_{n_{x-1}} \\ 
y_{3_{x}}=\alpha _{31}y_{1_{x-1}}+\alpha _{32}y_{2_{x-1}}+\alpha
_{33}y_{3_{x-1}}+\ldots +\alpha _{3n}y_{n_{x-1}} \\ 
\cdot \text{\qquad }\cdot \text{\qquad }\cdot \text{\qquad }\cdot \text{%
\qquad }\cdot \text{\qquad }\cdot \text{\qquad }\cdot \text{\qquad }\cdot 
\text{\qquad }\cdot \\ 
y_{n_{x}}=\alpha _{n1}y_{1_{x-1}}+\alpha _{n2}y_{2_{x-1}}+\alpha
_{n3}y_{3_{x-1}}+\ldots +\alpha _{nn}y_{n_{x-1}}.%
\end{array}%
\right.  \label{recf15}
\end{equation}

We present equations (\ref{recf15}) in the matrix-vectorial form as 
\begin{equation}
Y_{x}=AY_{x-1}  \label{recf17}
\end{equation}%
where 
\begin{equation*}
A=\left( 
\begin{array}{ccccc}
\alpha _{11} & \alpha _{12} & \alpha _{13} & \ldots & \alpha _{1n} \\ 
\alpha _{21} & \alpha _{22} & \alpha _{23} & \ldots & \alpha _{2n} \\ 
\alpha _{31} & \alpha _{32} & \alpha _{33} & \ldots & \alpha _{3n} \\ 
\cdot & \cdot & \cdot & \cdot & \cdot \\ 
\alpha _{n1} & \alpha _{n2} & \alpha _{n3} & \ldots & \alpha _{nn}%
\end{array}%
\right)
\end{equation*}
and 
\begin{equation*}
Y=\left( 
\begin{array}{c}
y_{1} \\ 
y_{2} \\ 
y_{3} \\ 
\cdot \\ 
y_{n}%
\end{array}%
\right) .
\end{equation*}

We will use the following basic notions.

The square matrix 
\begin{equation*}
I=\left( 
\begin{array}{ccccc}
1 & 0 & 0 & \ldots & 0 \\ 
0 & 1 & 0 & \ldots & 0 \\ 
0 & 0 & 1 & \ldots & 0 \\ 
\cdot & \cdot & \cdot & \cdot & \cdot \\ 
0 & 0 & 0 & \ldots & 1%
\end{array}%
\right)
\end{equation*}
is called the \textit{unit matrix}. If $A$ is a square matrix of order $n$
and $I$ is a unit matrix of order $n$, the $\lambda $-matrix 
\begin{equation*}
A-\lambda I=\left( 
\begin{array}{ccccc}
\alpha _{11}-\lambda & \alpha _{12} & \alpha _{13} & \ldots & \alpha _{1n}
\\ 
\alpha _{21} & \alpha _{22}-\lambda & \alpha _{23} & \ldots & \alpha _{2n}
\\ 
\alpha _{31} & \alpha _{32} & \alpha _{33}-\lambda & \ldots & \alpha _{3n}
\\ 
\cdot & \cdot & \cdot & \cdot & \cdot \\ 
\alpha _{n1} & \alpha _{n2} & \alpha _{n3} & \ldots & \alpha _{nn}-\lambda%
\end{array}%
\right)
\end{equation*}
is called the \textit{characteristic matrix of} $A$ and the determinant of $%
A $ is called the \textit{characteristic polynomial of} $A$.

By the \textit{Hamilton-Cayley theorem} \cite{Blo}, \cite{BoD}, if $A$ is a
square matrix and $\phi (\lambda )$ its characteristic polynomial, then 
\begin{equation*}
\phi (A)=0.
\end{equation*}
That is, if 
\begin{equation*}
\phi (\lambda )=u_{0}\lambda ^{n}+u_{1}\lambda ^{n-1}+u_{2}\lambda
^{n-2}+\ldots +u_{n-1}\lambda +u_{n}\text{ ,}
\end{equation*}
then 
\begin{equation*}
u_{0}A^{n}+u_{1}A^{n-1}+u_{2}A^{n-2}+\ldots +u_{n-1}A+u_{n}I=0.
\end{equation*}
As follows from the definition of the characteristic polynomial, $%
u_{0}=\left( -1\right) ^{n}$. Hence, 
\begin{equation}
A^{n}=\beta _{1}A^{n-1}+\beta _{2}A^{n-2}+\ldots +\beta _{n-1}A+\beta _{n}I
\label{recf16}
\end{equation}%
where $\beta _{j}=\left( -1\right) ^{n+1}u_{j}$ $(j=1,2,\ldots ,n)$. We
multiply the left and the right parts of (\ref{recf6}) by vector $Y_{x-n}$
and get: 
\begin{equation}
A^{n}Y_{x-n}=\beta _{1}A^{n-1}Y_{x-n}+\beta _{2}A^{n-2}Y_{x-n}+\ldots +\beta
_{n-1}AY_{x-n}+\beta _{n}Y_{x-n}.  \label{recf18}
\end{equation}%
On the other hand, iterating recurrence (\ref{recf17}) gives 
\begin{equation*}
Y_{x}=AY_{x-1}=A^{2}Y_{x-2}=\ldots =A^{n-1}Y_{x-n+1}=A^{n}Y_{x-n}\text{.}
\end{equation*}
After substituting these results in (\ref{recf18}) we have: 
\begin{equation*}
Y_{x}=\beta _{1}Y_{x-1}+\beta _{2}Y_{x-2}+\ldots +\beta
_{n-1}Y_{x-n+1}+\beta _{n}Y_{x-n}\text{.}
\end{equation*}
That is, 
\begin{equation}
\begin{array}{l}
y_{1_{x}}=\beta _{1}y_{1_{x-1}}+\beta _{2}y_{1_{x-2}}+\ldots +\beta
_{n-1}y_{1_{x-n+1}}+\beta _{n}y_{1_{x-n}} \\ 
y_{2_{x}}=\beta _{1}y_{2_{x-1}}+\beta _{2}y_{2_{x-2}}+\ldots +\beta
_{n-1}y_{2_{x-n+1}}+\beta _{n}y_{2_{x-n}} \\ 
y_{3_{x}}=\beta _{1}y_{3_{x-1}}+\beta _{2}y_{3_{x-2}}+\ldots +\beta
_{n-1}y_{3_{x-n+1}}+\beta _{n}y_{3_{x-n}} \\ 
\cdot \text{\qquad }\cdot \text{\qquad }\cdot \text{\qquad }\cdot \text{%
\qquad }\cdot \text{\qquad }\cdot \text{\qquad }\cdot \text{\qquad }\cdot 
\text{\qquad }\cdot \\ 
y_{n_{x}}=\beta _{1}y_{n_{x-1}}+\beta _{2}y_{n_{x-2}}+\ldots +\beta
_{n-1}y_{n_{x-n+1}}+\beta _{n}y_{n_{x-n}}%
\end{array}
\label{recf25}
\end{equation}%
where a coefficient $\beta _{j}$ $(j=1,2,\ldots ,n)$ equals a coefficient $%
u_{j}$ near a term $\lambda ^{n-j}$ in a characteristic polynomial of matrix 
$A$ for odd $n$ and $-u_{j}$ for even $n$.

For example, for $n=2$, when we have two the following recurrences: 
\begin{equation*}
\left\{ 
\begin{array}{l}
a_{x}=\alpha _{11}a_{x-1}+\alpha _{12}b_{x-1} \\ 
b_{x}=\alpha _{21}a_{x-1}+\alpha _{22}b_{x-1},%
\end{array}%
\right.
\end{equation*}
the result is 
\begin{equation}
\begin{array}{l}
a_{x}=(\alpha _{11}+\alpha _{22})a_{x-1}+(\alpha _{12}\alpha _{21}-\alpha
_{11}\alpha _{22})a_{x-2} \\ 
b_{x}=(\alpha _{11}+\alpha _{22})b_{x-1}+(\alpha _{12}\alpha _{21}-\alpha
_{11}\alpha _{22})b_{x-2}.%
\end{array}
\label{recf29}
\end{equation}

Now, move on to recurrences with absolute terms (\ref{recf14}). There are $%
n+1$ coefficients in each recurrence. We add the trivial $n+1$ equation to (%
\ref{recf14}) and get: 
\begin{equation}
\left\{ 
\begin{array}{l}
y_{1_{x}}=\alpha _{11}y_{1_{x-1}}+\alpha _{12}y_{2_{x-1}}+\alpha
_{13}y_{3_{x-1}}+\ldots +\alpha _{1n}y_{n_{x-1}}+\alpha _{1} \\ 
y_{2_{x}}=\alpha _{21}y_{1_{x-1}}+\alpha _{22}y_{2_{x-1}}+\alpha
_{23}y_{3_{x-1}}+\ldots +\alpha _{2n}y_{n_{x-1}}+\alpha _{2} \\ 
y_{3_{x}}=\alpha _{31}y_{1_{x-1}}+\alpha _{32}y_{2_{x-1}}+\alpha
_{33}y_{3_{x-1}}+\ldots +\alpha _{3n}y_{n_{x-1}}+\alpha _{3} \\ 
\cdot \text{\qquad }\cdot \text{\qquad }\cdot \text{\qquad }\cdot \text{%
\qquad }\cdot \text{\qquad }\cdot \text{\qquad }\cdot \text{\qquad }\cdot 
\text{\qquad }\cdot \text{\qquad }\cdot \\ 
y_{n_{x}}=\alpha _{n1}y_{1_{x-1}}+\alpha _{n2}y_{2_{x-1}}+\alpha
_{n3}y_{3_{x-1}}+\ldots +\alpha _{nn}y_{n_{x-1}}+\alpha _{n} \\ 
1\text{\quad }=0\text{\qquad \quad }+0\text{\qquad \quad }+0\text{\qquad }%
\quad +\ldots +0\text{\qquad }\quad +1.%
\end{array}%
\right.  \label{recf19}
\end{equation}%
Now, matrix $A^{*}$ of coefficients looks as 
\begin{equation*}
A^{*}=\left( 
\begin{array}{cccccc}
\alpha _{11} & \alpha _{12} & \alpha _{13} & \ldots & \alpha _{1n} & \alpha
_{1} \\ 
\alpha _{21} & \alpha _{22} & \alpha _{23} & \ldots & \alpha _{2n} & \alpha
_{2} \\ 
\alpha _{31} & \alpha _{32} & \alpha _{33} & \ldots & \alpha _{3n} & \alpha
_{3} \\ 
\cdot & \cdot & \cdot & \cdot & \cdot & \cdot \\ 
\alpha _{n1} & \alpha _{n2} & \alpha _{n3} & \ldots & \alpha _{nn} & \alpha
_{n} \\ 
0 & 0 & 0 & \ldots & 0 & 1%
\end{array}%
\right)
\end{equation*}
and vector $Y^{*}$ of variables is presented as 
\begin{equation*}
Y^{*}=\left( 
\begin{array}{c}
y_{1} \\ 
y_{2} \\ 
y_{3} \\ 
\cdot \\ 
y_{n} \\ 
1%
\end{array}%
\right) .
\end{equation*}
In such a case, equations (\ref{recf19}) are presented in the
matrix-vectorial form as 
\begin{equation}
Y_{x}^{*}=A^{*}Y_{x-1}^{*}.  \label{recf22}
\end{equation}

The characteristic polynomial of matrix $A^{*}$ looks as 
\begin{eqnarray}
\phi ^{*}(\lambda ) &=&\left| 
\begin{array}{cccccc}
\alpha _{11}-\lambda & \alpha _{12} & \alpha _{13} & \ldots & \alpha _{1n} & 
\alpha _{1} \\ 
\alpha _{21} & \alpha _{22}-\lambda & \alpha _{23} & \ldots & \alpha _{2n} & 
\alpha _{2} \\ 
\alpha _{31} & \alpha _{32} & \alpha _{33}-\lambda & \ldots & \alpha _{3n} & 
\alpha _{3} \\ 
\cdot & \cdot & \cdot & \cdot & \cdot & \cdot \\ 
\alpha _{n1} & \alpha _{n2} & \alpha _{n3} & \ldots & \alpha _{nn}-\lambda & 
\alpha _{n} \\ 
0 & 0 & 0 & \ldots & 0 & 1-\lambda%
\end{array}%
\right|  \notag \\
&&  \label{recf20} \\
&=&\left( -1\right) ^{2n+2}\left( 1-\lambda \right) \left| 
\begin{array}{ccccc}
\alpha _{11}-\lambda & \alpha _{12} & \alpha _{13} & \ldots & \alpha _{1n}
\\ 
\alpha _{21} & \alpha _{22}-\lambda & \alpha _{23} & \ldots & \alpha _{2n}
\\ 
\alpha _{31} & \alpha _{32} & \alpha _{33}-\lambda & \ldots & \alpha _{3n}
\\ 
\cdot & \cdot & \cdot & \cdot & \cdot \\ 
\alpha _{n1} & \alpha _{n2} & \alpha _{n3} & \ldots & \alpha _{nn}-\lambda%
\end{array}%
\right| \qquad \qquad  \notag \\
&=&\left( 1-\lambda \right) \phi (\lambda )  \notag
\end{eqnarray}%
(we apply an \textit{expansion of determinant by minors} \cite{Blo}).

By the Hamilton-Cayley theorem , if 
\begin{equation*}
\phi ^{*}(\lambda )=u_{0}^{*}\lambda ^{n+1}+u_{1}^{*}\lambda
^{n}+u_{2}^{*}\lambda ^{n-1}+\ldots +u_{n}^{*}\lambda +u_{n+1}^{*}\text{ ,}
\end{equation*}
then 
\begin{equation*}
u_{0}^{*}\left( A^{*}\right) ^{n+1}+u_{1}^{*}\left( A^{*}\right)
^{n}+u_{2}^{*}\left( A^{*}\right) ^{n-1}+\ldots
+u_{n}^{*}A^{*}+u_{n+1}^{*}I=0.
\end{equation*}
As follows from \ref{recf20}, $u_{0}^{*}=\left( -1\right) ^{n+1}$. Hence, 
\begin{equation}
\left( A^{*}\right) ^{n+1}=\beta _{1}^{*}\left( A^{*}\right) ^{n}+\beta
_{2}^{*}\left( A^{*}\right) ^{n-1}+\ldots +\beta _{n}^{*}A^{*}+\beta
_{n+1}^{*}I  \label{recf21}
\end{equation}%
where $\beta _{j}^{*}=\left( -1\right) ^{n}u_{j}^{*}$ $(j=1,2,\ldots ,n,n+1)$%
. We multiply the left and the right parts of (\ref{recf21}) by vector $%
Y_{x-n-1}^{*}$ and get: 
\begin{eqnarray}
\left( A^{*}\right) ^{n+1}Y_{x-n-1}^{*} &=&\beta _{1}^{*}\left( A^{*}\right)
^{n}Y_{x-n-1}^{*}+\beta _{2}^{*}\left( A^{*}\right)
^{n-1}Y_{x-n-1}^{*}+\ldots  \notag \\
&&+\beta _{n}^{*}A^{*}Y_{x-n-1}^{*}+\beta _{n+1}^{*}Y_{x-n-1}^{*}.
\label{recf23}
\end{eqnarray}%
Iterating recurrence \ref{recf22} gives 
\begin{equation*}
Y_{x}^{*}=A^{*}Y_{x-1}^{*}=\left( A^{*}\right) ^{2}Y_{x-2}^{*}=\ldots
=\left( A^{*}\right) ^{n}Y_{x-n}^{*}=\left( A^{*}\right) ^{n+1}Y_{x-n-1}^{*}
\end{equation*}
and after substituting these results in (\ref{recf23}) we have: 
\begin{equation*}
Y_{x}^{*}=\beta _{1}^{*}Y_{x-1}^{*}+\beta _{2}^{*}Y_{x-2}^{*}+\ldots +\beta
_{n}^{*}Y_{x-n}^{*}+\beta _{n+1}^{*}Y_{x-n-1}^{*}\text{.}
\end{equation*}
That is, 
\begin{equation}
\begin{array}{l}
y_{1_{x}}=\beta _{1}^{*}y_{1_{x-1}}+\beta _{2}^{*}y_{1_{x-2}}+\ldots +\beta
_{n}^{*}y_{1_{x-n}}+\beta _{n+1}^{*}y_{1_{x-n-1}} \\ 
y_{2_{x}}=\beta _{1}^{*}y_{2_{x-1}}+\beta _{2}^{*}y_{2_{x-2}}+\ldots +\beta
_{n}^{*}y_{2_{x-n}}+\beta _{n+1}^{*}y_{2_{x-n-1}} \\ 
y_{3_{x}}=\beta _{1}^{*}y_{3_{x-1}}+\beta _{2}^{*}y_{3_{x-2}}+\ldots +\beta
_{n}^{*}y_{3_{x-n}}+\beta _{n+1}^{*}y_{3_{x-n-1}} \\ 
\cdot \text{\qquad }\cdot \text{\qquad }\cdot \text{\qquad }\cdot \text{%
\qquad }\cdot \text{\qquad }\cdot \text{\qquad }\cdot \text{\qquad }\cdot 
\text{\qquad }\cdot \\ 
y_{n_{x}}=\beta _{1}^{*}y_{n_{x-1}}+\beta _{2}^{*}y_{n_{x-2}}+\ldots +\beta
_{n}^{*}y_{n_{x-n}}+\beta _{n+1}^{*}y_{n_{x-n-1}} \\ 
1=\beta _{1}^{*}+\beta _{2}^{*}+\ldots +\beta _{n}^{*}+\beta _{n+1}^{*}%
\end{array}
\label{recf24}
\end{equation}%
where a coefficient $\beta _{j}^{*}$ $(j=1,2,\ldots ,n,n+1)$ equals a
coefficient $u_{j}^{*}$ near a term $\lambda ^{n+1-j}$ in a characteristic
polynomial of matrix $A^{*}$ for even $n$ and $-u_{j}$ for odd $n$.

For example, for $n=1$, we have a recurrence 
\begin{equation*}
a_{x}=\alpha _{11}a_{x-1}+\alpha _{1}
\end{equation*}
that can be presented as 
\begin{equation*}
a_{x}=(\alpha _{11}+1)a_{x-1}-\alpha _{11}a_{x-2}\text{ ;}
\end{equation*}
for $n=2$, we have two the following recurrences: 
\begin{equation}
\left\{ 
\begin{array}{l}
a_{x}=\alpha _{11}a_{x-1}+\alpha _{12}b_{x-1}+\alpha _{1} \\ 
b_{x}=\alpha _{21}a_{x-1}+\alpha _{22}b_{x-1}+\alpha _{2}%
\end{array}%
\right.  \label{recf26}
\end{equation}%
and the result is 
\begin{equation*}
\begin{array}{l}
a_{x}=(\alpha _{11}+\alpha _{22}+1)a_{x-1}+(\alpha _{12}\alpha _{21}-\alpha
_{11}\alpha _{22}-\alpha _{11}-\alpha _{22})a_{x-2}+ \\ 
\qquad \,(\alpha _{11}\alpha _{22}-\alpha _{12}\alpha _{21})a_{x-3}\smallskip
\\ 
b_{x}=(\alpha _{11}+\alpha _{22}+1)b_{x-1}+(\alpha _{12}\alpha _{21}-\alpha
_{11}\alpha _{22}-\alpha _{11}-\alpha _{22})b_{x-2}+ \\ 
\qquad \,(\alpha _{11}\alpha _{22}-\alpha _{12}\alpha _{21})b_{x-3}.%
\end{array}%
\end{equation*}

Thus, we expressed each variable $y_{i}$ $\left( i=1,2,\ldots ,n\right) $
from (\ref{recf14}) in its $n+1$ previous values. A coefficient $\beta
_{j}^{*}$ near a value $y_{i_{x-j}}$ $(j=1,2,\ldots ,n,n+1)$ does not depend
on $i$, i.e., all recurrences (\ref{recf24}) have the same structure.
Absolute terms $\alpha _{1},\alpha _{2},\ldots ,\alpha _{n}$ do not appear
explicitly in (\ref{recf24}). They are accounted for implicitly, by addition
of a supplementary $n+1$ previous value of $y_{i}$ in the expressions
(compare (\ref{recf24}) with ((\ref{recf25})).

Nevertheless, it is also of interest to express variables of (\ref{recf14})
directly using absolute terms. Such a presentation is more elegant since it
contains explicitly all the information about initial recurrences.
Specifically, it can be deduced for $n=2$ and $n=3$ through a number of
transformations.

Consider recurrences (\ref{recf26}). As follows from the first equation, 
\begin{eqnarray}
\alpha _{12}b_{x-1} &=&a_{x}-\alpha _{11}a_{x-1}-\alpha _{1}\Leftrightarrow 
\notag \\
b_{x-1} &=&\frac{a_{x}}{\alpha _{12}}-\frac{\alpha _{11}}{\alpha _{12}}%
a_{x-1}-\frac{\alpha _{1}}{\alpha _{12}}\Leftrightarrow  \label{recf27} \\
b_{x} &=&\frac{a_{x+1}}{\alpha _{12}}-\frac{\alpha _{11}}{\alpha _{12}}a_{x}-%
\frac{\alpha _{1}}{\alpha _{12}}.  \label{recf28}
\end{eqnarray}%
We substitute (\ref{recf27}) and (\ref{recf28}) into the second equation (%
\ref{recf26}) and get: 
\begin{eqnarray*}
\frac{a_{x+1}}{\alpha _{12}}-\frac{\alpha _{11}}{\alpha _{12}}a_{x}-\frac{%
\alpha _{1}}{\alpha _{12}} &=&\alpha _{21}a_{x-1}+\alpha _{22}\left( \frac{%
a_{x}}{\alpha _{12}}-\frac{\alpha _{11}}{\alpha _{12}}a_{x-1}-\frac{\alpha
_{1}}{\alpha _{12}}\right) +\alpha _{2}\Leftrightarrow \\
\frac{a_{x}}{\alpha _{12}}-\frac{\alpha _{11}}{\alpha _{12}}a_{x-1}-\frac{%
\alpha _{1}}{\alpha _{12}} &=&\alpha _{21}a_{x-2}+\alpha _{22}\left( \frac{%
a_{x-1}}{\alpha _{12}}-\frac{\alpha _{11}}{\alpha _{12}}a_{x-2}-\frac{\alpha
_{1}}{\alpha _{12}}\right) +\alpha _{2}\Leftrightarrow \\
a_{x} &=&\left( \alpha _{11}+\alpha _{22}\right) a_{x-1}+(\alpha _{12}\alpha
_{21}-\alpha _{11}\alpha _{22})a_{x-2}+ \\
&&\alpha _{1}\left( 1-\alpha _{22}\right) +\alpha _{2}\alpha _{12}.
\end{eqnarray*}%
The expression for $b_{x}$ is derived in the same way and, finally, we have: 
\begin{equation}
\begin{array}{l}
a_{x}=(\alpha _{11}+\alpha _{22})a_{x-1}+(\alpha _{12}\alpha _{21}-\alpha
_{11}\alpha _{22})a_{x-2}+ \\ 
\qquad \,\alpha _{1}\left( 1-\alpha _{22}\right) +\alpha _{2}\alpha
_{12}\smallskip \\ 
b_{x}=(\alpha _{11}+\alpha _{22})b_{x-1}+(\alpha _{12}\alpha _{21}-\alpha
_{11}\alpha _{22})b_{x-2}+ \\ 
\qquad \,\alpha _{2}\left( 1-\alpha _{11}\right) +\alpha _{1}\alpha _{21}%
\end{array}
\label{recf30}
\end{equation}%
(compare with (\ref{recf29})).

For $n=3$, recurrences (\ref{recf14}) can be presented as three following
equations: 
\begin{equation}
\left\{ 
\begin{array}{l}
a_{x}=\alpha _{11}a_{x-1}+\alpha _{12}b_{x-1}+\alpha _{13}c_{x-1}+\alpha _{1}
\\ 
b_{x}=\alpha _{21}a_{x-1}+\alpha _{22}b_{x-1}+\alpha _{23}c_{x-1}+\alpha _{2}
\\ 
c_{x}=\alpha _{31}a_{x-1}+\alpha _{32}b_{x-1}+\alpha _{33}c_{x-1}+\alpha
_{3}.%
\end{array}%
\right.  \label{recf5}
\end{equation}%
The derivation is similar to the derivation for $n=2$ and the final result
is the following: 
\begin{equation}
\begin{array}{l}
a_{x}=(\alpha _{11}+\alpha _{22}+\alpha _{33})a_{x-1}+ \\ 
\qquad \,(\alpha _{12}\alpha _{21}+\alpha _{13}\alpha _{31}+\alpha
_{23}\alpha _{32}-\alpha _{11}\alpha _{22}-\alpha _{11}\alpha _{33}-\alpha
_{22}\alpha _{33})a_{x-2}+ \\ 
\qquad \,(\alpha _{11}\alpha _{22}\alpha _{33}+\alpha _{12}\alpha
_{23}\alpha _{31}+\alpha _{21}\alpha _{13}\alpha _{32}-\alpha _{12}\alpha
_{21}\alpha _{33}-\alpha _{13}\alpha _{31}\alpha _{22}- \\ 
\qquad \,\alpha _{23}\alpha _{32}\alpha _{11})a_{x-3}+ \\ 
\qquad \,\alpha _{1}\left( 1-\alpha _{22}-\alpha _{33}\right) +\alpha
_{2}\alpha _{12}+\alpha _{3}\alpha _{13}+\alpha _{1}\left( \alpha
_{22}\alpha _{33}-\alpha _{23}\alpha _{32}\right) + \\ 
\qquad \,\alpha _{2}\left( \alpha _{13}\alpha _{32}-\alpha _{33}\alpha
_{12}\right) +\alpha _{3}\left( \alpha _{12}\alpha _{23}-\alpha _{22}\alpha
_{13}\right) \smallskip \\ 
b_{x}=(\alpha _{11}+\alpha _{22}+\alpha _{33})b_{x-1}+ \\ 
\qquad \,(\alpha _{12}\alpha _{21}+\alpha _{13}\alpha _{31}+\alpha
_{23}\alpha _{32}-\alpha _{11}\alpha _{22}-\alpha _{11}\alpha _{33}-\alpha
_{22}\alpha _{33})b_{x-2}+ \\ 
\qquad \,(\alpha _{11}\alpha _{22}\alpha _{33}+\alpha _{12}\alpha
_{23}\alpha _{31}+\alpha _{21}\alpha _{13}\alpha _{32}-\alpha _{12}\alpha
_{21}\alpha _{33}-\alpha _{13}\alpha _{31}\alpha _{22}- \\ 
\qquad \,\alpha _{23}\alpha _{32}\alpha _{11})b_{x-3}+ \\ 
\qquad \,\alpha _{2}\left( 1-\alpha _{11}-\alpha _{33}\right) +\alpha
_{1}\alpha _{21}+\alpha _{3}\alpha _{23}+\alpha _{2}\left( \alpha
_{11}\alpha _{33}-\alpha _{13}\alpha _{31}\right) + \\ 
\qquad \,\alpha _{3}\left( \alpha _{21}\alpha _{13}-\alpha _{11}\alpha
_{23}\right) +\alpha _{1}\left( \alpha _{23}\alpha _{31}-\alpha _{33}\alpha
_{21}\right) \smallskip \\ 
c_{x}=(\alpha _{11}+\alpha _{22}+\alpha _{33})c_{x-1}+ \\ 
\qquad \,(\alpha _{12}\alpha _{21}+\alpha _{13}\alpha _{31}+\alpha
_{23}\alpha _{32}-\alpha _{11}\alpha _{22}-\alpha _{11}\alpha _{33}-\alpha
_{22}\alpha _{33})c_{x-2}+ \\ 
\qquad \,(\alpha _{11}\alpha _{22}\alpha _{33}+\alpha _{12}\alpha
_{23}\alpha _{31}+\alpha _{21}\alpha _{13}\alpha _{32}-\alpha _{12}\alpha
_{21}\alpha _{33}-\alpha _{13}\alpha _{31}\alpha _{22}- \\ 
\qquad \,\alpha _{23}\alpha _{32}\alpha _{11})c_{x-3}+ \\ 
\qquad \,\alpha _{3}\left( 1-\alpha _{11}-\alpha _{22}\right) +\alpha
_{1}\alpha _{31}+\alpha _{2}\alpha _{32}+\alpha _{3}\left( \alpha
_{11}\alpha _{22}-\alpha _{12}\alpha _{21}\right) + \\ 
\qquad \,\alpha _{1}\left( \alpha _{32}\alpha _{21}-\alpha _{22}\alpha
_{31}\right) +\alpha _{2}\left( \alpha _{31}\alpha _{12}-\alpha _{11}\alpha
_{32}\right) .%
\end{array}%
\qquad  \label{recf31}
\end{equation}

Hence, each variable $y_{i}$ $\left( i=1,2,\ldots ,n\right) $ from (\ref%
{recf14}) is expressed in its $n$ previous values for $n=2,3$ if absolute
terms appear explicitly in expressions. As follows from (\ref{recf30}) and (%
\ref{recf31}) the structure of absolute terms in these expressions depends
on $i$.

\section{A Solution of Matrix Recurrences of Order Three by their
Decomposition to Matrix Recurrences of Order Two\label{sec_2_3_sim_rec}}

We intend to solve three simultaneous recurrences (\ref{recf5}), where $x$
is a current number of variables $a$, $b$, and $c$ in their sequences
generated by means of (\ref{recf5}) so that $a_{0}$, $b_{0}$, and $c_{0}$
are initial values of $a$, $b$, and $c$, respectively. That is, we intend to
express $a_{x}$, $b_{x}$, and $c_{x}$ of (\ref{recf5}) directly in $x$.

It would be possible to solve this problem by the general method described
in the previous section and then to apply the technique of solving regular
recurrences from \cite{Ros}. However, under certain conditions, there is a
simpler way of arriving at a solution. This way is decomposition of (\ref%
{recf5}) to matrix recurrences of order two.

Matrix recurrences of order two can also be solved by the general method.
However, we use another way that seems to be more efficient under the
special conditions.

\begin{lemma}
\label{lem_2rec}If 
\begin{equation*}
\left\{ 
\begin{array}{l}
a_{x}=\alpha _{11}a_{x-1}+\alpha _{12}b_{x-1}+\alpha _{1} \\ 
b_{x}=\alpha _{21}a_{x-1}+\alpha _{22}b_{x-1}+\alpha _{2}%
\end{array}%
\right.
\end{equation*}%
and 
\begin{equation}
\alpha _{11}+\alpha _{12}=\alpha _{21}+\alpha _{22}\text{ ,}  \label{recf6}
\end{equation}%
then

1. When $\alpha _{12}\neq -\alpha _{21}$, $\alpha _{11}+\alpha _{12}\neq 1$, 
$\alpha _{11}-\alpha _{21}\neq 1$%
\begin{eqnarray*}
a_{x} &=&\left( \alpha _{11}+\alpha _{12}\right) ^{x}a_{0}+\alpha
_{12}\left( b_{0}-a_{0}\right) \frac{\left( \alpha _{11}+\alpha _{12}\right)
^{x}-\left( \alpha _{11}-\alpha _{21}\right) ^{x}}{\alpha _{12}+\alpha _{21}}%
+ \\
&&\alpha _{1}\frac{\left( \alpha _{11}+\alpha _{12}\right) ^{x}-1}{\alpha
_{11}+\alpha _{12}-1}+\frac{\alpha _{12}\left( \alpha _{2}-\alpha
_{1}\right) }{\alpha _{12}+\alpha _{21}}\times \\
&&\left( \frac{\left( \alpha _{11}+\alpha _{12}\right) ^{x}-\alpha
_{11}-\alpha _{12}}{\alpha _{11}+\alpha _{12}-1}-\frac{\left( \alpha
_{11}-\alpha _{21}\right) ^{x}-\alpha _{11}+\alpha _{21}}{\alpha
_{11}-\alpha _{21}-1}\right) \\
b_{x} &=&\left( \alpha _{11}+\alpha _{12}\right) ^{x}b_{0}+\alpha
_{21}\left( a_{0}-b_{0}\right) \frac{\left( \alpha _{11}+\alpha _{12}\right)
^{x}-\left( \alpha _{11}-\alpha _{21}\right) ^{x}}{\alpha _{12}+\alpha _{21}}%
+ \\
&&\alpha _{2}\frac{\left( \alpha _{11}+\alpha _{12}\right) ^{x}-1}{\alpha
_{11}+\alpha _{12}-1}+\frac{\alpha _{21}\left( \alpha _{1}-\alpha
_{2}\right) }{\alpha _{12}+\alpha _{21}}\times \\
&&\left( \frac{\left( \alpha _{11}+\alpha _{12}\right) ^{x}-\alpha
_{11}-\alpha _{12}}{\alpha _{11}+\alpha _{12}-1}-\frac{\left( \alpha
_{11}-\alpha _{21}\right) ^{x}-\alpha _{11}+\alpha _{21}}{\alpha
_{11}-\alpha _{21}-1}\right) \text{.}
\end{eqnarray*}

2. When $\alpha _{12}\neq -\alpha _{21}$, $\alpha _{11}+\alpha _{12}=1$ ($%
\alpha _{11}-\alpha _{21}\neq 1$) 
\begin{eqnarray*}
a_{x} &=&\left( \alpha _{11}+\alpha _{12}\right) ^{x}a_{0}+\alpha
_{12}\left( b_{0}-a_{0}\right) \frac{\left( \alpha _{11}+\alpha _{12}\right)
^{x}-\left( \alpha _{11}-\alpha _{21}\right) ^{x}}{\alpha _{12}+\alpha _{21}}%
+ \\
&&\alpha _{1}x+\frac{\alpha _{12}\left( \alpha _{2}-\alpha _{1}\right) }{%
\alpha _{11}-\alpha _{21}-1}\left( \frac{\left( \alpha _{11}-\alpha
_{21}\right) ^{x}-\alpha _{11}+\alpha _{21}}{\alpha _{11}-\alpha _{21}-1}%
-x+1\right) \\
b_{x} &=&\left( \alpha _{11}+\alpha _{12}\right) ^{x}b_{0}+\alpha
_{21}\left( a_{0}-b_{0}\right) \frac{\left( \alpha _{11}+\alpha _{12}\right)
^{x}-\left( \alpha _{11}-\alpha _{21}\right) ^{x}}{\alpha _{12}+\alpha _{21}}%
+ \\
&&\alpha _{2}x+\frac{\alpha _{21}\left( \alpha _{1}-\alpha _{2}\right) }{%
\alpha _{11}-\alpha _{21}-1}\left( \frac{\left( \alpha _{11}-\alpha
_{21}\right) ^{x}-\alpha _{11}+\alpha _{21}}{\alpha _{11}-\alpha _{21}-1}%
-x+1\right) \text{.}
\end{eqnarray*}

3. When $\alpha _{12}\neq -\alpha _{21}$, $\alpha _{11}-\alpha _{21}=1$ ($%
\alpha _{11}+\alpha _{12}\neq 1$) 
\begin{eqnarray*}
a_{x} &=&\left( \alpha _{11}+\alpha _{12}\right) ^{x}a_{0}+\alpha
_{12}\left( b_{0}-a_{0}\right) \frac{\left( \alpha _{11}+\alpha _{12}\right)
^{x}-\left( \alpha _{11}-\alpha _{21}\right) ^{x}}{\alpha _{12}+\alpha _{21}}%
+ \\
&&\alpha _{1}\frac{\left( \alpha _{11}+\alpha _{12}\right) ^{x}-1}{\alpha
_{11}+\alpha _{12}-1}+\frac{\alpha _{12}\left( \alpha _{2}-\alpha
_{1}\right) }{\alpha _{11}+\alpha _{12}-1}\left( \frac{\left( \alpha
_{11}+\alpha _{12}\right) ^{x}-\alpha _{11}-\alpha _{12}}{\alpha
_{11}+\alpha _{12}-1}-x+1\right) \\
b_{x} &=&\left( \alpha _{11}+\alpha _{12}\right) ^{x}b_{0}+\alpha
_{21}\left( a_{0}-b_{0}\right) \frac{\left( \alpha _{11}+\alpha _{12}\right)
^{x}-\left( \alpha _{11}-\alpha _{21}\right) ^{x}}{\alpha _{12}+\alpha _{21}}%
+ \\
&&\alpha _{2}\frac{\left( \alpha _{11}+\alpha _{12}\right) ^{x}-1}{\alpha
_{11}+\alpha _{12}-1}+\frac{\alpha _{21}\left( \alpha _{1}-\alpha
_{2}\right) }{\alpha _{11}+\alpha _{12}-1}\left( \frac{\left( \alpha
_{11}+\alpha _{12}\right) ^{x}-\alpha _{11}-\alpha _{12}}{\alpha
_{11}+\alpha _{12}-1}-x+1\right) \text{.}
\end{eqnarray*}

4. When $\alpha _{12}=-\alpha _{21}$, $\alpha _{11}+\alpha _{12}\neq 1$ ($%
\alpha _{11}-\alpha _{21}\neq 1$) 
\begin{eqnarray*}
a_{x} &=&\left( \alpha _{11}+\alpha _{12}\right) ^{x-1}\left( \left( \alpha
_{11}+\alpha _{12}\right) a_{0}+\alpha _{12}\left( b_{0}-a_{0}\right)
x\right) + \\
&&\alpha _{1}\frac{\left( \alpha _{11}+\alpha _{12}\right) ^{x}-1}{\alpha
_{11}+\alpha _{12}-1}+\frac{\alpha _{12}\left( \alpha _{2}-\alpha
_{1}\right) }{\left( \alpha _{11}+\alpha _{12}-1\right) ^{2}}\times \\
&&\left( \left( x-1\right) \left( \alpha _{11}+\alpha _{12}\right)
^{x}-x\left( \alpha _{11}+\alpha _{12}\right) ^{x-1}+1\right) \\
b_{x} &=&\left( \alpha _{11}+\alpha _{12}\right) ^{x-1}\left( \left( \alpha
_{11}+\alpha _{12}\right) b_{0}+\alpha _{21}\left( a_{0}-b_{0}\right)
x\right) + \\
&&\alpha _{2}\frac{\left( \alpha _{11}+\alpha _{12}\right) ^{x}-1}{\alpha
_{11}+\alpha _{12}-1}+\frac{\alpha _{21}\left( \alpha _{1}-\alpha
_{2}\right) }{\left( \alpha _{11}+\alpha _{12}-1\right) ^{2}}\times \\
&&\left( \left( x-1\right) \left( \alpha _{11}+\alpha _{12}\right)
^{x}-x\left( \alpha _{11}+\alpha _{12}\right) ^{x-1}+1\right) \text{.}
\end{eqnarray*}

5. When $\alpha _{12}=-\alpha _{21}$, $\alpha _{11}+\alpha _{12}=1$ ($\alpha
_{11}-\alpha _{21}=1$) 
\begin{eqnarray*}
a_{x} &=&\left( \alpha _{11}+\alpha _{12}\right) ^{x-1}\left( \left( \alpha
_{11}+\alpha _{12}\right) a_{0}+\alpha _{12}\left( b_{0}-a_{0}\right)
x\right) + \\
&&\alpha _{1}x+\alpha _{12}\left( \alpha _{2}-\alpha _{1}\right) \frac{%
x\left( x-1\right) }{2} \\
b_{x} &=&\left( \alpha _{11}+\alpha _{12}\right) ^{x-1}\left( \left( \alpha
_{11}+\alpha _{12}\right) b_{0}+\alpha _{21}\left( a_{0}-b_{0}\right)
x\right) + \\
&&\alpha _{2}x+\alpha _{21}\left( \alpha _{1}-\alpha _{2}\right) \frac{%
x\left( x-1\right) }{2}\text{.}
\end{eqnarray*}
\end{lemma}

\proof%
1. Denote $\Delta _{x}=b_{x}-a_{x}$, $S=\alpha _{11}+\alpha _{12}$, $%
D=\alpha _{11}-\alpha _{21}$, $\delta =\alpha _{2}-\alpha _{1}$. In such a
case, 
\begin{eqnarray*}
a_{x} &=&\alpha _{11}a_{x-1}+\alpha _{12}\left( a_{x-1}+\Delta _{x-1}\right)
+\alpha _{1} \\
&=&\left( \alpha _{11}+\alpha _{12}\right) a_{x-1}+\alpha _{12}\Delta
_{x-1}+\alpha _{1} \\
&=&Sa_{x-1}+\alpha _{12}\Delta _{x-1}+\alpha _{1}
\end{eqnarray*}%
As follows from (\ref{recf6}) 
\begin{equation*}
\alpha _{11}-\alpha _{21}=\alpha _{22}-\alpha _{12}\text{.}
\end{equation*}%
Therefore, 
\begin{equation*}
b_{x}-a_{x}=\left( \alpha _{11}-\alpha _{21}\right) (b_{x-1}-a_{x-1})+\alpha
_{2}-\alpha _{1}
\end{equation*}%
or 
\begin{equation*}
\Delta _{x}=D\Delta _{x-1}+\delta \text{.}
\end{equation*}%
Hence, we have two simultaneous recurrences: 
\begin{equation}
\left\{ 
\begin{array}{l}
a_{x}=Sa_{x-1}+\alpha _{12}\Delta _{x-1}+\alpha _{1} \\ 
\Delta _{x}=D\Delta _{x-1}+\delta \text{.}%
\end{array}%
\right.  \label{recf7}
\end{equation}%
Based on (\ref{recf7}) we get: 
\begin{eqnarray*}
a_{x} &=&Sa_{x-1}+\alpha _{12}\Delta _{x-1}+\alpha _{1} \\
&=&S\left( Sa_{x-2}+\alpha _{12}\Delta _{x-2}+\alpha _{1}\right) +\alpha
_{12}\left( D\Delta _{x-2}+\delta \right) +\alpha _{1} \\
&=&S^{2}a_{x-2}+\alpha _{12}\left( S+D\right) \Delta _{x-2}+\left(
S+1\right) \alpha _{1}+\alpha _{12}\delta \\
&=&S^{2}\left( Sa_{x-3}+\alpha _{12}\Delta _{x-3}+\alpha _{1}\right) +\alpha
_{12}\left( S+D\right) \left( D\Delta _{x-3}+\delta \right) +\left(
S+1\right) \alpha _{1}+\alpha _{12}\delta \\
&=&S^{3}a_{x-3}+\alpha _{12}\left( S^{2}+SD+D^{2}\right) \Delta
_{x-3}+\left( S^{2}+S+1\right) \alpha _{1}+\alpha _{12}\left( S+D+1\right)
\delta \\
&=&S^{4}a_{x-4}+\alpha _{12}\left( S^{3}+S^{2}D+SD^{2}+D^{3}\right) \Delta
_{x-4}+ \\
&&\left( S^{3}+S^{2}+S+1\right) \alpha _{1}+\alpha _{12}\left(
S^{2}+SD+D^{2}+S+D+1\right) \delta \\
&=&\ldots =S^{x}a_{0}+\alpha _{12}\Delta _{0}\underset{i=0}{\overset{x-1}{%
\sum }}S^{i}D^{x-1-i}+\alpha _{1}\underset{i=0}{\overset{x-1}{\sum }}%
S^{i}+\alpha _{12}\delta \underset{j=0}{\overset{x-2}{\sum }}\underset{i=0}{%
\overset{j}{\sum }}S^{i}D^{j-i} \\
&=&S^{x}a_{0}+\alpha _{12}\Delta _{0}D^{x-1}\underset{i=0}{\overset{x-1}{%
\sum }}\left( \frac{S}{D}\right) ^{i}+\alpha _{1}\frac{S^{x}-1}{S-1}+\alpha
_{12}\delta \underset{j=0}{\overset{x-2}{\sum }}D^{j}\underset{i=0}{\overset{%
j}{\sum }}\left( \frac{S}{D}\right) ^{i} \\
&=&S^{x}a_{0}+\alpha _{12}\Delta _{0}D^{x-1}\frac{\left( \frac{S}{D}\right)
^{x}-1}{\frac{S}{D}-1}+\alpha _{1}\frac{S^{x}-1}{S-1}+\alpha _{12}\delta 
\underset{j=0}{\overset{x-2}{\sum }}D^{j}\frac{\left( \frac{S}{D}\right)
^{j+1}-1}{\frac{S}{D}-1} \\
&=&S^{x}a_{0}+\alpha _{12}\Delta _{0}\frac{S^{x}-D^{x}}{S-D}+\alpha _{1}%
\frac{S^{x}-1}{S-1}+\alpha _{12}\delta \underset{j=0}{\overset{x-2}{\sum }}%
\frac{S^{j+1}-D^{j+1}}{S-D} \\
&=&S^{x}a_{0}+\alpha _{12}\Delta _{0}\frac{S^{x}-D^{x}}{S-D}+\alpha _{1}%
\frac{S^{x}-1}{S-1}+\frac{\alpha _{12}\delta }{S-D}\left( \overset{x-2}{%
\underset{j=0}{\sum }}S^{j+1}-\overset{x-2}{\underset{j=0}{\sum }}%
D^{j+1}\right) \\
&=&S^{x}a_{0}+\alpha _{12}\Delta _{0}\frac{S^{x}-D^{x}}{S-D}+\alpha _{1}%
\frac{S^{x}-1}{S-1}+\frac{\alpha _{12}\delta }{S-D}\left( \frac{S\left(
S^{x-1}-1\right) }{S-1}-\frac{D\left( D^{x-1}-1\right) }{D-1}\right) \\
&=&S^{x}a_{0}+\alpha _{12}\Delta _{0}\frac{S^{x}-D^{x}}{S-D}+\alpha _{1}%
\frac{S^{x}-1}{S-1}+\frac{\alpha _{12}\delta }{S-D}\left( \frac{S^{x}-S}{S-1}%
-\frac{D^{x}-D}{D-1}\right) \\
&=&\left( \alpha _{11}+\alpha _{12}\right) ^{x}a_{0}+\alpha _{12}\left(
b_{0}-a_{0}\right) \frac{\left( \alpha _{11}+\alpha _{12}\right) ^{x}-\left(
\alpha _{11}-\alpha _{21}\right) ^{x}}{\alpha _{12}+\alpha _{21}}+ \\
&&\alpha _{1}\frac{\left( \alpha _{11}+\alpha _{12}\right) ^{x}-1}{\alpha
_{11}+\alpha _{12}-1}+\frac{\alpha _{12}\left( \alpha _{2}-\alpha
_{1}\right) }{\alpha _{12}+\alpha _{21}}\times \\
&&\left( \frac{\left( \alpha _{11}+\alpha _{12}\right) ^{x}-\alpha
_{11}-\alpha _{12}}{\alpha _{11}+\alpha _{12}-1}-\frac{\left( \alpha
_{11}-\alpha _{21}\right) ^{x}-\alpha _{11}+\alpha _{21}}{\alpha
_{11}-\alpha _{21}-1}\right) \text{.}
\end{eqnarray*}%
We used that 
\begin{equation*}
S-D=\alpha _{11}+\alpha _{12}-\alpha _{11}+\alpha _{21}=\alpha _{12}+\alpha
_{21}\text{.}
\end{equation*}%
The result for $b_{x}$ can be derived in the same way.

Statements 2, 3, 4, 5 are proven similarly. 
\endproof%

\begin{lemma}
\label{lem_3rec}If 
\begin{equation}
\left\{ 
\begin{array}{l}
a_{x}=\alpha _{11}a_{x-1}+\alpha _{12}b_{x-1}+\alpha _{13}c_{x-1}+\alpha _{1}
\\ 
b_{x}=\alpha _{21}a_{x-1}+\alpha _{22}b_{x-1}+\alpha _{23}c_{x-1}+\alpha _{2}
\\ 
c_{x}=\alpha _{31}a_{x-1}+\alpha _{32}b_{x-1}+\alpha _{33}c_{x-1}+\alpha _{3}%
\end{array}%
\right.  \label{recf8}
\end{equation}%
and the following conditions hold:

1. $\alpha _{11}+\alpha _{12}+\alpha _{13}=\alpha _{21}+\alpha _{22}+\alpha
_{23}=\alpha _{31}+\alpha _{32}+\alpha _{33}$

2. $\exists $ $w_{1},w_{2}$, $w_{1}+w_{2}=1$, that $\forall $ $x$, $%
b_{x}=w_{1}a_{x}+w_{2}c_{x}$,\newline
then three simultaneous recurrences (\ref{recf8}) can be presented by means
of representations of $c$ through $a$ and $b$, $b$ through $a$ and $c$, and $%
a$ through $b$ and $c$ as the three pairs of the following simultaneous
recurrences, respectively: 
\begin{equation}
\left\{ 
\begin{array}{l}
a_{x}=\alpha _{11}^{^{\prime }}a_{x-1}+\alpha _{12}^{^{\prime
}}b_{x-1}+\alpha _{1} \\ 
b_{x}=\alpha _{21}^{^{\prime }}a_{x-1}+\alpha _{22}^{^{\prime
}}b_{x-1}+\alpha _{2}%
\end{array}%
\right.  \label{recf9}
\end{equation}%
where $\alpha _{11}^{^{\prime }}=\alpha _{11}-\frac{w_{1}}{w_{2}}\alpha
_{13} $, $\alpha _{12}^{^{\prime }}=\alpha _{12}+\frac{1}{w_{2}}\alpha _{13}$%
, $\alpha _{21}^{^{\prime }}=\alpha _{21}-\frac{w_{1}}{w_{2}}\alpha _{23}$, $%
\alpha _{22}^{^{\prime }}=\alpha _{22}+\frac{1}{w_{2}}\alpha _{23}$; 
\begin{equation}
\left\{ 
\begin{array}{l}
a_{x}=\alpha _{11}^{^{\prime }}a_{x-1}+\alpha _{12}^{^{\prime
}}c_{x-1}+\alpha _{1} \\ 
c_{x}=\alpha _{21}^{^{\prime }}a_{x-1}+\alpha _{22}^{^{\prime
}}c_{x-1}+\alpha _{3}%
\end{array}%
\right.  \label{recf10}
\end{equation}%
where $\alpha _{11}^{^{\prime }}=\alpha _{11}+w_{1}\alpha _{12}$, $\alpha
_{12}^{^{\prime }}=w_{2}\alpha _{12}+\alpha _{13}$, $\alpha _{21}^{^{\prime
}}=\alpha _{31}+w_{1}\alpha _{32}$, $\alpha _{22}^{^{\prime }}=w_{2}\alpha
_{32}+\alpha _{33}$; 
\begin{equation}
\left\{ 
\begin{array}{l}
b_{x}=\alpha _{11}^{^{\prime }}b_{x-1}+\alpha _{12}^{^{\prime
}}c_{x-1}+\alpha _{2} \\ 
c_{x}=\alpha _{21}^{^{\prime }}b_{x-1}+\alpha _{22}^{^{\prime
}}c_{x-1}+\alpha _{3}%
\end{array}%
\right.  \label{recf11}
\end{equation}%
where $\alpha _{11}^{^{\prime }}=\frac{1}{w_{1}}\alpha _{21}+\alpha _{22}$, $%
\alpha _{12}^{^{\prime }}=-\frac{w_{2}}{w_{1}}\alpha _{21}+\alpha _{23}$, $%
\alpha _{21}^{^{\prime }}=\frac{1}{w_{1}}\alpha _{31}+\alpha _{32}$, $\alpha
_{22}^{^{\prime }}=-\frac{w_{2}}{w1}\alpha _{31}+\alpha _{33}$ and for all
these pairs of simultaneous recurrences 
\begin{equation*}
\alpha _{11}^{^{\prime }}+\alpha _{12}^{^{\prime }}=\alpha _{21}^{^{\prime
}}+\alpha _{22}^{^{\prime }}\text{.}
\end{equation*}
\end{lemma}

\proof%
Since $b_{x}=w_{1}a_{x}+w_{2}c_{x}$ then $c_{x}=-\frac{w_{1}}{w_{2}}a_{x}+%
\frac{1}{w_{2}}b_{x}$. We substitute $c$ in (\ref{recf8}) and get: 
\begin{eqnarray*}
a_{x} &=&\alpha _{11}a_{x-1}+\alpha _{12}b_{x-1}+\alpha _{13}c_{x-1}+\alpha
_{1} \\
&=&\alpha _{11}a_{x-1}+\alpha _{12}b_{x-1}+\alpha _{13}\left( \frac{1}{w_{2}}%
b_{x-1}-\frac{w_{1}}{w_{2}}a_{x-1}\right) +\alpha _{1} \\
&=&\left( \alpha _{11}-\frac{w_{1}}{w_{2}}\alpha _{13}\right) a_{x-1}+\left(
\alpha _{12}+\frac{1}{w_{2}}\alpha _{13}\right) b_{x-1}+\alpha _{1} \\
b_{x} &=&\alpha _{21}a_{x-1}+\alpha _{22}b_{x-1}+\alpha _{23}c_{x-1}+\alpha
_{2} \\
&=&\alpha _{21}a_{x-1}+\alpha _{22}b_{x-1}+\alpha _{23}\left( \frac{1}{w_{2}}%
b_{x-1}-\frac{w_{1}}{w_{2}}a_{x-1}\right) +\alpha _{2} \\
&=&\left( \alpha _{21}-\frac{w_{1}}{w_{2}}\alpha _{23}\right) a_{x-1}+\left(
\alpha _{22}+\frac{1}{w_{2}}\alpha _{23}\right) b_{x-1}+\alpha _{2}\text{.}
\end{eqnarray*}%
Here 
\begin{eqnarray*}
\alpha _{11}^{^{\prime }}+\alpha _{12}^{^{\prime }} &=&\alpha _{11}-\frac{%
w_{1}}{w_{2}}\alpha _{13}+\alpha _{12}+\frac{1}{w_{2}}\alpha _{13} \\
&=&\alpha _{11}+\alpha _{12}+\alpha _{13} \\
\alpha _{21}^{^{\prime }}+\alpha _{22}^{^{\prime }} &=&\alpha _{21}-\frac{%
w_{1}}{w_{2}}\alpha _{23}+\alpha _{22}+\frac{1}{w_{2}}\alpha _{23} \\
&=&\alpha _{21}+\alpha _{22}+\alpha _{23}\text{.}
\end{eqnarray*}%
Since $\alpha _{11}+\alpha _{12}+\alpha _{13}=\alpha _{21}+\alpha
_{22}+\alpha _{23}$ then $\alpha _{11}^{^{\prime }}+\alpha _{12}^{^{\prime
}}=\alpha _{21}^{^{\prime }}+\alpha _{22}^{^{\prime }}$.

We substitute $b$ in (\ref{recf8}) and get: 
\begin{eqnarray*}
a_{x} &=&\alpha _{11}a_{x-1}+\alpha _{12}b_{x-1}+\alpha _{13}c_{x-1}+\alpha
_{1} \\
&=&\alpha _{11}a_{x-1}+\alpha _{12}\left( w_{1}a_{x-1}+w_{2}c_{x-1}\right)
+\alpha _{13}c_{x-1}+\alpha _{1} \\
&=&\left( \alpha _{11}+w_{1}\alpha _{12}\right) a_{x-1}+\left( w_{2}\alpha
_{12}+\alpha _{13}\right) c_{x-1}+\alpha _{1} \\
c_{x} &=&\alpha _{31}a_{x-1}+\alpha _{32}b_{x-1}+\alpha _{33}c_{x-1}+\alpha
_{3} \\
&=&\alpha _{31}a_{x-1}+\alpha _{32}\left( w_{1}a_{x-1}+w_{2}c_{x-1}\right)
+\alpha _{33}c_{x-1}+\alpha _{3} \\
&=&\left( \alpha _{31}+w_{1}\alpha _{32}\right) a_{x-1}+\left( w_{2}\alpha
_{32}+\alpha _{33}\right) c_{x-1}+\alpha _{3}\text{.}
\end{eqnarray*}%
Here 
\begin{eqnarray*}
\alpha _{11}^{^{\prime }}+\alpha _{12}^{^{\prime }} &=&\alpha
_{11}+w_{1}\alpha _{12}+w_{2}\alpha _{12}+\alpha _{13} \\
&=&\alpha _{11}+\alpha _{12}+\alpha _{13} \\
\alpha _{21}^{^{\prime }}+\alpha _{22}^{^{\prime }} &=&\alpha
_{31}+w_{1}\alpha _{32}+w_{2}\alpha _{32}+\alpha _{33} \\
&=&\alpha _{31}+\alpha _{32}+\alpha _{33}\text{.}
\end{eqnarray*}%
Since $\alpha _{11}+\alpha _{12}+\alpha _{13}=\alpha _{31}+\alpha
_{32}+\alpha _{33}$ then $\alpha _{11}^{^{\prime }}+\alpha _{12}^{^{\prime
}}=\alpha _{21}^{^{\prime }}+\alpha _{22}^{^{\prime }}$.

The pair of simultaneous recurrences for $b$ and $c$ and the equality $%
\alpha _{11}^{^{\prime }}+\alpha _{12}^{^{\prime }}=\alpha _{21}^{^{\prime
}}+\alpha _{22}^{^{\prime }}$ for this case are derived in the same way. 
\endproof%

\begin{remark}
Denote $w_{1}^{^{\prime }}=-\frac{w_{1}}{w_{2}}$, $w_{2}^{^{\prime }}=\frac{1%
}{w_{2}}$, $w_{1}^{^{\prime \prime }}=\frac{1}{w_{1}}$, $w_{2}^{^{\prime
\prime }}=-\frac{w_{2}}{w_{1}}$. In such a case, $c=w_{1}^{^{\prime
}}a+w_{2}^{^{\prime }}b$ and $a=w_{1}^{^{\prime \prime }}b+w_{2}^{^{\prime
\prime }}c$, where $w_{1}^{^{\prime }}+w_{2}^{^{\prime }}=w_{1}^{^{\prime
\prime }}+w_{2}^{^{\prime \prime }}=1$. Hence, special conditions
accompanying (\ref{recf8}) hold for any combination of recurrences. For this
reason, it would be enough to prove the equality $\alpha _{11}^{^{\prime
}}+\alpha _{12}^{^{\prime }}=\alpha _{21}^{^{\prime }}+\alpha
_{22}^{^{\prime }}$ for one pair of simultaneous recurrences in Lemma \ref%
{lem_3rec}. The equality holds automatically for the other two pairs in this
case.
\end{remark}

\begin{theorem}
\label{th_3rec}If 
\begin{equation*}
\left\{ 
\begin{array}{l}
a_{x}=\alpha _{11}a_{x-1}+\alpha _{12}b_{x-1}+\alpha _{13}c_{x-1}+\alpha _{1}
\\ 
b_{x}=\alpha _{21}a_{x-1}+\alpha _{22}b_{x-1}+\alpha _{23}c_{x-1}+\alpha _{2}
\\ 
c_{x}=\alpha _{31}a_{x-1}+\alpha _{32}b_{x-1}+\alpha _{33}c_{x-1}+\alpha _{3}%
\end{array}%
\right.
\end{equation*}%
and the following conditions hold:

1. $\alpha _{11}+\alpha _{12}+\alpha _{13}=\alpha _{21}+\alpha _{22}+\alpha
_{23}=\alpha _{31}+\alpha _{32}+\alpha _{33}$

2. $\exists $ $w_{1},w_{2}$, $w_{1}+w_{2}=1$, that $\forall $ $x$, $%
b_{x}=w_{1}a_{x}+w_{2}c_{x}$,\newline
then for $C_{1}=$ $\alpha _{11}+\alpha _{12}+\alpha _{13}$, $C_{2}=\alpha
_{12}+\frac{1}{w_{2}}\alpha _{13}$, $C_{3}=\alpha _{11}-\frac{w_{1}}{w_{2}}%
\alpha _{13}-\alpha _{21}+\frac{w_{1}}{w_{2}}\alpha _{23}$, $C_{4}=\alpha
_{12}+\frac{1}{w_{2}}\alpha _{13}+\alpha _{21}-\frac{w_{1}}{w_{2}}\alpha
_{23}$, $C_{5}=\alpha _{21}-\frac{w_{1}}{w_{2}}\alpha _{23}$:

1. When $\alpha _{12}+\frac{1}{w_{2}}\alpha _{13}+\alpha _{21}-\frac{w_{1}}{%
w_{2}}\alpha _{23}\neq 0$, $\alpha _{11}+\alpha _{12}+\alpha _{13}\neq 1$, $%
\alpha _{11}-\frac{w_{1}}{w_{2}}\alpha _{13}-\alpha _{21}+\frac{w_{1}}{w_{2}}%
\alpha _{23}\neq 1$%
\begin{eqnarray*}
a_{x} &=&\left( C_{1}\right) ^{x}a_{0}+C_{2}\frac{\left( C_{1}\right)
^{x}-\left( C_{3}\right) ^{x}}{C_{4}}\left( b_{0}-a_{0}\right) +\alpha _{1}%
\frac{\left( C_{1}\right) ^{x}-1}{C_{1}-1}+ \\
&&\frac{\left( \alpha _{2}-\alpha _{1}\right) C_{2}}{C_{4}}\left( \frac{%
\left( C_{1}\right) ^{x}-C_{1}}{C_{1}-1}-\frac{\left( C_{3}\right) ^{x}-C_{3}%
}{C_{3}-1}\right) \\
b_{x} &=&\left( C_{1}\right) ^{x}b_{0}+C_{5}\frac{\left( C_{1}\right)
^{x}-\left( C_{3}\right) ^{x}}{C_{4}}\left( a_{0}-b_{0}\right) +\alpha _{2}%
\frac{\left( C_{1}\right) ^{x}-1}{C_{1}-1}+ \\
&&\frac{\left( \alpha _{1}-\alpha _{2}\right) C_{5}}{C_{4}}\left( \frac{%
\left( C_{1}\right) ^{x}-C_{1}}{C_{1}-1}-\frac{\left( C_{3}\right) ^{x}-C_{3}%
}{C_{3}-1}\right) \\
c_{x} &=&-\frac{w_{1}}{w_{2}}a_{x}+\frac{1}{w_{2}}b_{x}\text{.}
\end{eqnarray*}

2. When $\alpha _{12}+\frac{1}{w_{2}}\alpha _{13}+\alpha _{21}-\frac{w_{1}}{%
w_{2}}\alpha _{23}\neq 0$, $\alpha _{11}+\alpha _{12}+\alpha _{13}=1$ ($%
\alpha _{11}-\frac{w_{1}}{w_{2}}\alpha _{13}-\alpha _{21}+\frac{w_{1}}{w_{2}}%
\alpha _{23}\neq 1$) 
\begin{eqnarray*}
a_{x} &=&\left( C_{1}\right) ^{x}a_{0}+C_{2}\frac{\left( C_{1}\right)
^{x}-\left( C_{3}\right) ^{x}}{C_{4}}\left( b_{0}-a_{0}\right) +\alpha _{1}x+
\\
&&\frac{\left( \alpha _{2}-\alpha _{1}\right) C_{2}}{C_{3}-1}\left( \frac{%
\left( C_{3}\right) ^{x}-C_{3}}{C_{3}-1}-x+1\right) \\
b_{x} &=&\left( C_{1}\right) ^{x}b_{0}+C_{5}\frac{\left( C_{1}\right)
^{x}-\left( C_{3}\right) ^{x}}{C_{4}}\left( a_{0}-b_{0}\right) +\alpha _{2}x+
\\
&&\frac{\left( \alpha _{1}-\alpha _{2}\right) C_{5}}{C_{3}-1}\left( \frac{%
\left( C_{3}\right) ^{x}-C_{3}}{C_{3}-1}-x+1\right) \\
c_{x} &=&-\frac{w_{1}}{w_{2}}a_{x}+\frac{1}{w_{2}}b_{x}\text{.}
\end{eqnarray*}

3. When $\alpha _{12}+\frac{1}{w_{2}}\alpha _{13}+\alpha _{21}-\frac{w_{1}}{%
w_{2}}\alpha _{23}\neq 0$, $\alpha _{11}-\frac{w_{1}}{w_{2}}\alpha
_{13}-\alpha _{21}+\frac{w_{1}}{w_{2}}\alpha _{23}=1$ ($\alpha _{11}+\alpha
_{12}+\alpha _{13}\neq 1$) 
\begin{eqnarray*}
a_{x} &=&\left( C_{1}\right) ^{x}a_{0}+C_{2}\frac{\left( C_{1}\right)
^{x}-\left( C_{3}\right) ^{x}}{C_{4}}\left( b_{0}-a_{0}\right) +\alpha _{1}%
\frac{\left( C_{1}\right) ^{x}-1}{C_{1}-1}+ \\
&&\frac{\left( \alpha _{2}-\alpha _{1}\right) C_{2}}{C_{1}-1}\left( \frac{%
\left( C_{1}\right) ^{x}-C_{1}}{C_{1}-1}-x+1\right) \\
b_{x} &=&\left( C_{1}\right) ^{x}b_{0}+C_{5}\frac{\left( C_{1}\right)
^{x}-\left( C_{3}\right) ^{x}}{C_{4}}\left( a_{0}-b_{0}\right) +\alpha _{2}%
\frac{\left( C_{1}\right) ^{x}-1}{C_{1}-1}+ \\
&&\frac{\left( \alpha _{1}-\alpha _{2}\right) C_{5}}{C_{1}-1}\left( \frac{%
\left( C_{1}\right) ^{x}-C_{1}}{C_{1}-1}-x+1\right) \\
c_{x} &=&-\frac{w_{1}}{w_{2}}a_{x}+\frac{1}{w_{2}}b_{x}\text{.}
\end{eqnarray*}

4. When $\alpha _{12}+\frac{1}{w_{2}}\alpha _{13}+\alpha _{21}-\frac{w_{1}}{%
w_{2}}\alpha _{23}=0$, $\alpha _{11}+\alpha _{12}+\alpha _{13}\neq 1$ ($%
\alpha _{11}-\frac{w_{1}}{w_{2}}\alpha _{13}-\alpha _{21}+\frac{w_{1}}{w_{2}}%
\alpha _{23}\neq 1$) 
\begin{eqnarray*}
a_{x} &=&\left( C_{1}\right) ^{x-1}\left( C_{1}a_{0}+C_{2}\left(
b_{0}-a_{0}\right) x\right) +\alpha _{1}\frac{\left( C_{1}\right) ^{x}-1}{%
C_{1}-1}+ \\
&&\frac{\left( \alpha _{2}-\alpha _{1}\right) C_{2}}{\left( C_{1}-1\right)
^{2}}\left( \left( x-1\right) \left( C_{1}\right) ^{x}-x\left( C_{1}\right)
^{x-1}+1\right) \\
b_{x} &=&\left( C_{1}\right) ^{x-1}\left( C_{1}b_{0}+C_{5}\left(
a_{0}-b_{0}\right) x\right) +\alpha _{2}\frac{\left( C_{1}\right) ^{x}-1}{%
C_{1}-1}+ \\
&&\frac{\left( \alpha _{1}-\alpha _{2}\right) C_{5}}{\left( C_{1}-1\right)
^{2}}\left( \left( x-1\right) \left( C_{1}\right) ^{x}-x\left( C_{1}\right)
^{x-1}+1\right) \\
c_{x} &=&-\frac{w_{1}}{w_{2}}a_{x}+\frac{1}{w_{2}}b_{x}\text{.}
\end{eqnarray*}

5. When $\alpha _{12}+\frac{1}{w_{2}}\alpha _{13}+\alpha _{21}-\frac{w_{1}}{%
w_{2}}\alpha _{23}=0$, $\alpha _{11}+\alpha _{12}+\alpha _{13}=1$ ($\alpha
_{11}-\frac{w_{1}}{w_{2}}\alpha _{13}-\alpha _{21}+\frac{w_{1}}{w_{2}}\alpha
_{23}=1$) 
\begin{eqnarray*}
a_{x} &=&\left( C_{1}\right) ^{x-1}\left( C_{1}a_{0}+C_{2}\left(
b_{0}-a_{0}\right) x\right) +\alpha _{1}x+C_{2}\left( \alpha _{2}-\alpha
_{1}\right) \frac{x\left( x-1\right) }{2} \\
b_{x} &=&\left( C_{1}\right) ^{x-1}\left( C_{1}b_{0}+C_{5}\left(
a_{0}-b_{0}\right) x\right) +\alpha _{2}x+C_{5}\left( \alpha _{1}-\alpha
_{2}\right) \frac{x\left( x-1\right) }{2} \\
c_{x} &=&-\frac{w_{1}}{w_{2}}a_{x}+\frac{1}{w_{2}}b_{x}\text{.}
\end{eqnarray*}
\end{theorem}

\proof%
As follows from Lemma \ref{lem_3rec}, three simultaneous recurrences (\ref%
{recf8}) can be reduced, specifically, to the pair of simultaneous
recurrences (\ref{recf9}). Based on Lemma \ref{lem_2rec} and after
corresponding substitutions we obtain the expressions for $a_{x}$ and $b_{x}$%
. The result for $c_{x}$ is obtained automatically. 
\endproof%

\begin{remark}
It is not the only way to determine $a_{x}$, $b_{x}$, and $c_{x}$ that is
presented in Theorem \ref{th_3rec}. Each of recurrent variables $a_{x}$, $%
b_{x}$, and $c_{x}$ can be determined through two of three pairs of
simultaneous recurrences (\ref{recf9} -- \ref{recf11}).
\end{remark}

\begin{remark}
In principle, it could be sufficient to derive only one direct expression
(for $a_{x}$, or $b_{x}$, or $c_{x}$). For example, if the formula for $%
a_{x} $ is derived, then $b_{x}$ and $c_{x}$ can be expressed in $a_{x}$ and
in the difference between $b_{x}$ and $a_{x}$. Actually, the difference $%
b_{x}-a_{x}$ was obtained implicitly in the course of the proof of Lemma \ref%
{lem_2rec}. Based on the second recurrence in (\ref{recf7}) we have: 
\begin{equation*}
\Delta _{x}=D\Delta _{x-1}+\delta =D^{x}\Delta _{0}+\delta \frac{D^{x-1}-1}{%
D-1}
\end{equation*}%
for $D\neq 1$ and 
\begin{equation*}
\Delta _{x}=D\Delta _{x-1}+\delta =\Delta _{0}+x\delta
\end{equation*}%
for $D=1$. Hence, for (\ref{recf8}) and $C_{3}=\alpha _{11}-\frac{w_{1}}{%
w_{2}}\alpha _{13}-\alpha _{21}+\frac{w_{1}}{w_{2}}\alpha _{23}$ 
\begin{equation*}
b_{x}-a_{x}=\left( C_{3}\right) ^{x}\left( b_{0}-a_{0}\right) +\left( \alpha
_{2}-\alpha _{1}\right) \frac{\left( C_{3}\right) ^{x-1}-1}{C_{3}-1}
\end{equation*}%
when $C_{3}\neq 1$ and 
\begin{equation*}
b_{x}-a_{x}=b_{0}-a_{0}+x\left( \alpha _{2}-\alpha _{1}\right)
\end{equation*}%
when $C_{3}=1$. Relations between the differences are the following: 
\begin{equation*}
c_{x}-a_{x}=\frac{b_{x}-a_{x}}{w_{2}}=\frac{c_{x}-b_{x}}{w_{1}}\text{.}
\end{equation*}
\end{remark}

The equality $b_{x}=w_{1}a_{x}+w_{2}c_{x}$ does not automatically mean the
existence of the same proportion for coefficients ($\alpha _{21}=w_{1}\alpha
_{11}+w_{2}\alpha _{31}$, $\alpha _{22}=w_{1}\alpha _{12}+w_{2}\alpha _{32}$%
, $\alpha _{23}=w_{1}\alpha _{13}+w_{2}\alpha _{33}$, $\alpha
_{2}=w_{1}\alpha _{1}+w_{3}\alpha _{3}$, as in (\ref{recf32})). In the
special case, the linear combination $b_{x}=w_{1}a_{x}+w_{2}c_{x}$ can be
provided by the corresponding proportion for initial values of $a$, $b$, and 
$c$ without observing the same proportion for coefficients. For example, it
can be shown that in (\ref{recf33}) $b_{x}=$ $\frac{13}{19}a_{x}+$ $\frac{6}{%
19}c_{x}$ for $a_{0}=41$, $b_{0}=47$, $c_{0}=60$ although this proportion
does not take place for coefficients. Now, suppose that we have stronger
conditions when also the proportion between coefficients takes place.

\begin{lemma}
\label{lem_pro_co}If

1. $\alpha _{11}+\alpha _{12}+\alpha _{13}=\alpha _{31}+\alpha _{32}+\alpha
_{33}$

and

2. $\exists $ $w_{1},w_{2}$, $w_{1}+w_{2}=1$, that $\alpha _{21}=w_{1}\alpha
_{11}+w_{2}\alpha _{31}$, $\alpha _{22}=w_{1}\alpha _{12}+w_{2}\alpha _{32}$%
, $\alpha _{23}=w_{1}\alpha _{13}+w_{2}\alpha _{33}$,

then 
\begin{equation}
\alpha _{21}+\alpha _{22}+\alpha _{23}=\alpha _{11}+\alpha _{12}+\alpha
_{13}=\alpha _{31}+\alpha _{32}+\alpha _{33}.  \label{recf13}
\end{equation}
\end{lemma}

\proof%
\begin{eqnarray*}
\alpha _{21}+\alpha _{22}+\alpha _{23} &=&w_{1}\alpha _{11}+w_{2}\alpha
_{31}+w_{1}\alpha _{12}+w_{2}\alpha _{32}+w_{1}\alpha _{13}+w_{2}\alpha _{33}
\\
&=&w_{1}(\alpha _{11}+\alpha _{12}+\alpha _{13})+w_{2}(\alpha _{31}+\alpha
_{32}+\alpha _{33}) \\
&=&\left( w_{1}+w_{2}\right) \left( \alpha _{11}+\alpha _{12}+\alpha
_{13}\right) \\
&=&\alpha _{11}+\alpha _{12}+\alpha _{13}\text{.}
\end{eqnarray*}%
\endproof%
\medskip

Hence, the equality such as (\ref{recf13}) is a redundant condition if the
condition $b_{x}=w_{1}a_{x}+w_{2}c_{x}$ is provided by the corresponding
proportion between coefficients. The equality of any two sums of
coefficients from three ones in (\ref{recf13}) is sufficient in this case.

Consider a situation with weaker conditions when it is only known that 
\begin{equation*}
\alpha _{11}+\alpha _{12}+\alpha _{13}=\alpha _{31}+\alpha _{32}+\alpha _{33}
\end{equation*}%
and $b_{x}=w_{1}a_{x}+w_{2}c_{x}$, where $w_{1}+w_{2}=1$. In this case, only
one pair of simultaneous recurrences adduced in Lemma \ref{lem_3rec} can be
generated. However, the solution of this pair of recurrences is sufficient
for finding $a_{x}$, $b_{x}$, and $c_{x}$. On the other hand, two other
pairs of simultaneous recurrences (not as in Lemma \ref{lem_3rec}) can be
generated in this case also. Indeed, 
\begin{eqnarray*}
b_{x} &=&w_{1}a_{x}+w_{2}c_{x} \\
&=&w_{1}(\alpha _{11}a_{x-1}+\alpha _{12}b_{x-1}+\alpha _{13}c_{x-1}+\alpha
_{1})+ \\
&&w_{2}(\alpha _{31}a_{x-1}+\alpha _{32}b_{x-1}+\alpha _{33}c_{x-1}+\alpha
_{3}) \\
&=&\left( w_{1}\alpha _{11}+w_{2}\alpha _{31}\right) a_{x-1}+\left(
w_{1}\alpha _{12}+w_{2}\alpha _{32}\right) b_{x-1}+\left( w_{1}\alpha
_{13}+w_{2}\alpha _{33}\right) c_{x-1}+ \\
&&w_{1}\alpha _{1}+w_{2}\alpha _{3} \\
&=&\alpha _{21}^{\ast }a_{x-1}+\alpha _{22}^{\ast }b_{x-1}+\alpha
_{23}^{\ast }c_{x-1}+\alpha _{2}^{\ast }\text{.}
\end{eqnarray*}%
where $\alpha _{21}^{\ast }=w_{1}\alpha _{11}+w_{2}\alpha _{31}$, $\alpha
_{22}^{\ast }=w_{1}\alpha _{12}+w_{2}\alpha _{32}$, $\alpha _{23}^{\ast
}=w_{1}\alpha _{13}+w_{2}\alpha _{33}$, $\alpha _{2}^{\ast }=w_{1}\alpha
_{1}+w_{3}\alpha _{3}$. Hence, the recurrence for $b_{x}$ in (\ref{recf8})
can be replaced so that we have the following three simultaneous
recurrences: 
\begin{equation}
\left\{ 
\begin{array}{l}
a_{x}=\alpha _{11}a_{x-1}+\alpha _{12}b_{x-1}+\alpha _{13}c_{x-1}+\alpha _{1}
\\ 
b_{x}=\alpha _{21}^{\ast }a_{x-1}+\alpha _{22}^{\ast }b_{x-1}+\alpha
_{23}^{\ast }c_{x-1}+\alpha _{2}^{\ast } \\ 
c_{x}=\alpha _{31}a_{x-1}+\alpha _{32}b_{x-1}+\alpha _{33}c_{x-1}+\alpha _{3}%
\text{ .}%
\end{array}%
\right.  \label{recf12}
\end{equation}%
By Lemma \ref{lem_pro_co}, these recurrences are of the same structure as (%
\ref{recf8}). For this reason, by Lemma \ref{lem_3rec}, (\ref{recf12}) can
be presented as three pairs of simultaneous recurrences. Thus, Theorem \ref%
{th_3rec} can be generalized as follows.

\begin{theorem}
\label{th_3rec_gen}If 
\begin{equation*}
\left\{ 
\begin{array}{l}
a_{x}=\alpha _{11}a_{x-1}+\alpha _{12}b_{x-1}+\alpha _{13}c_{x-1}+\alpha _{1}
\\ 
b_{x}=\alpha _{21}a_{x-1}+\alpha _{22}b_{x-1}+\alpha _{23}c_{x-1}+\alpha _{2}
\\ 
c_{x}=\alpha _{31}a_{x-1}+\alpha _{32}b_{x-1}+\alpha _{33}c_{x-1}+\alpha _{3}%
\end{array}%
\right.
\end{equation*}%
and the following conditions hold:

1. $\alpha _{11}+\alpha _{12}+\alpha _{13}=\alpha _{31}+\alpha _{32}+\alpha
_{33}$

2. $\exists $ $w_{1},w_{2}$, $w_{1}+w_{2}=1$, that $\forall $ $x$, $%
b_{x}=w_{1}a_{x}+w_{2}c_{x}$,\newline
then for $C_{1}=$ $\alpha _{11}+\alpha _{12}+\alpha _{13}$, $C_{2}=\alpha
_{12}+\frac{1}{w_{2}}\alpha _{13}$, $C_{3}=\alpha _{11}-\frac{w_{1}}{w_{2}}%
\alpha _{13}-\alpha _{21}^{*}+\frac{w_{1}}{w_{2}}\alpha _{23}^{*}$, $%
C_{4}=\alpha _{12}+\frac{1}{w_{2}}\alpha _{13}+\alpha _{21}^{*}-\frac{w_{1}}{%
w_{2}}\alpha _{23}^{*}$, $C_{5}=\alpha _{21}^{*}-\frac{w_{1}}{w_{2}}\alpha
_{23}^{*}$, $\alpha _{21}^{*}=w_{1}\alpha _{11}+w_{2}\alpha _{31}$, $\alpha
_{23}^{*}=w_{1}\alpha _{13}+w_{2}\alpha _{33}$, $\alpha _{2}^{*}=w_{1}\alpha
_{1}+w_{3}\alpha _{3}$:

1. When $\alpha _{12}+\frac{1}{w_{2}}\alpha _{13}+\alpha _{21}^{*}-\frac{%
w_{1}}{w_{2}}\alpha _{23}^{*}\neq 0$, $\alpha _{11}+\alpha _{12}+\alpha
_{13}\neq 1$, $\alpha _{11}-\frac{w_{1}}{w_{2}}\alpha _{13}-\alpha _{21}^{*}+%
\frac{w_{1}}{w_{2}}\alpha _{23}^{*}\neq 1$%
\begin{eqnarray*}
a_{x} &=&\left( C_{1}\right) ^{x}a_{0}+C_{2}\frac{\left( C_{1}\right)
^{x}-\left( C_{3}\right) ^{x}}{C_{4}}\left( b_{0}-a_{0}\right) +\alpha _{1}%
\frac{\left( C_{1}\right) ^{x}-1}{C_{1}-1}+ \\
&&\frac{\left( \alpha _{2}^{*}-\alpha _{1}\right) C_{2}}{C_{4}}\left( \frac{%
\left( C_{1}\right) ^{x}-C_{1}}{C_{1}-1}-\frac{\left( C_{3}\right) ^{x}-C_{3}%
}{C_{3}-1}\right) \\
b_{x} &=&\left( C_{1}\right) ^{x}b_{0}+C_{5}\frac{\left( C_{1}\right)
^{x}-\left( C_{3}\right) ^{x}}{C_{4}}\left( a_{0}-b_{0}\right) +\alpha
_{2}^{*}\frac{\left( C_{1}\right) ^{x}-1}{C_{1}-1}+ \\
&&\frac{\left( \alpha _{1}-\alpha _{2}^{*}\right) C_{5}}{C_{4}}\left( \frac{%
\left( C_{1}\right) ^{x}-C_{1}}{C_{1}-1}-\frac{\left( C_{3}\right) ^{x}-C_{3}%
}{C_{3}-1}\right) \\
c_{x} &=&-\frac{w_{1}}{w_{2}}a_{x}+\frac{1}{w_{2}}b_{x}\text{.}
\end{eqnarray*}

2. When $\alpha _{12}+\frac{1}{w_{2}}\alpha _{13}+\alpha _{21}^{*}-\frac{%
w_{1}}{w_{2}}\alpha _{23}^{*}\neq 0$, $\alpha _{11}+\alpha _{12}+\alpha
_{13}=1$ ($\alpha _{11}-\frac{w_{1}}{w_{2}}\alpha _{13}-\alpha _{21}^{*}+%
\frac{w_{1}}{w_{2}}\alpha _{23}^{*}\neq 1$) 
\begin{eqnarray*}
a_{x} &=&\left( C_{1}\right) ^{x}a_{0}+C_{2}\frac{\left( C_{1}\right)
^{x}-\left( C_{3}\right) ^{x}}{C_{4}}\left( b_{0}-a_{0}\right) +\alpha _{1}x+
\\
&&\frac{\left( \alpha _{2}^{*}-\alpha _{1}\right) C_{2}}{C_{3}-1}\left( 
\frac{\left( C_{3}\right) ^{x}-C_{3}}{C_{3}-1}-x+1\right) \\
b_{x} &=&\left( C_{1}\right) ^{x}b_{0}+C_{5}\frac{\left( C_{1}\right)
^{x}-\left( C_{3}\right) ^{x}}{C_{4}}\left( a_{0}-b_{0}\right) +\alpha
_{2}^{*}x+ \\
&&\frac{\left( \alpha _{1}-\alpha _{2}^{*}\right) C_{5}}{C_{3}-1}\left( 
\frac{\left( C_{3}\right) ^{x}-C_{3}}{C_{3}-1}-x+1\right) \\
c_{x} &=&-\frac{w_{1}}{w_{2}}a_{x}+\frac{1}{w_{2}}b_{x}\text{.}
\end{eqnarray*}

3. When $\alpha _{12}+\frac{1}{w_{2}}\alpha _{13}+\alpha _{21}^{*}-\frac{%
w_{1}}{w_{2}}\alpha _{23}^{*}\neq 0$, $\alpha _{11}-\frac{w_{1}}{w_{2}}%
\alpha _{13}-\alpha _{21}^{*}+\frac{w_{1}}{w_{2}}\alpha _{23}^{*}=1$ ($%
\alpha _{11}+\alpha _{12}+\alpha _{13}\neq 1$) 
\begin{eqnarray*}
a_{x} &=&\left( C_{1}\right) ^{x}a_{0}+C_{2}\frac{\left( C_{1}\right)
^{x}-\left( C_{3}\right) ^{x}}{C_{4}}\left( b_{0}-a_{0}\right) +\alpha _{1}%
\frac{\left( C_{1}\right) ^{x}-1}{C_{1}-1}+ \\
&&\frac{\left( \alpha _{2}^{*}-\alpha _{1}\right) C_{2}}{C_{1}-1}\left( 
\frac{\left( C_{1}\right) ^{x}-C_{1}}{C_{1}-1}-x+1\right) \\
b_{x} &=&\left( C_{1}\right) ^{x}b_{0}+C_{5}\frac{\left( C_{1}\right)
^{x}-\left( C_{3}\right) ^{x}}{C_{4}}\left( a_{0}-b_{0}\right) +\alpha
_{2}^{*}\frac{\left( C_{1}\right) ^{x}-1}{C_{1}-1}+ \\
&&\frac{\left( \alpha _{1}-\alpha _{2}^{*}\right) C_{5}}{C_{1}-1}\left( 
\frac{\left( C_{1}\right) ^{x}-C_{1}}{C_{1}-1}-x+1\right) \\
c_{x} &=&-\frac{w_{1}}{w_{2}}a_{x}+\frac{1}{w_{2}}b_{x}\text{.}
\end{eqnarray*}

4. When $\alpha _{12}+\frac{1}{w_{2}}\alpha _{13}+\alpha _{21}^{*}-\frac{%
w_{1}}{w_{2}}\alpha _{23}^{*}=0$, $\alpha _{11}+\alpha _{12}+\alpha
_{13}\neq 1$ ($\alpha _{11}-\frac{w_{1}}{w_{2}}\alpha _{13}-\alpha _{21}^{*}+%
\frac{w_{1}}{w_{2}}\alpha _{23}^{*}\neq 1$) 
\begin{eqnarray*}
a_{x} &=&\left( C_{1}\right) ^{x-1}\left( C_{1}a_{0}+C_{2}\left(
b_{0}-a_{0}\right) x\right) +\alpha _{1}\frac{\left( C_{1}\right) ^{x}-1}{%
C_{1}-1}+ \\
&&\frac{\left( \alpha _{2}^{*}-\alpha _{1}\right) C_{2}}{\left(
C_{1}-1\right) ^{2}}\left( \left( x-1\right) \left( C_{1}\right)
^{x}-x\left( C_{1}\right) ^{x-1}+1\right) \\
b_{x} &=&\left( C_{1}\right) ^{x-1}\left( C_{1}b_{0}+C_{5}\left(
a_{0}-b_{0}\right) x\right) +\alpha _{2}^{*}\frac{\left( C_{1}\right) ^{x}-1%
}{C_{1}-1}+ \\
&&\frac{\left( \alpha _{1}-\alpha _{2}^{*}\right) C_{5}}{\left(
C_{1}-1\right) ^{2}}\left( \left( x-1\right) \left( C_{1}\right)
^{x}-x\left( C_{1}\right) ^{x-1}+1\right) \\
c_{x} &=&-\frac{w_{1}}{w_{2}}a_{x}+\frac{1}{w_{2}}b_{x}\text{.}
\end{eqnarray*}

5. When $\alpha _{12}+\frac{1}{w_{2}}\alpha _{13}+\alpha _{21}^{*}-\frac{%
w_{1}}{w_{2}}\alpha _{23}^{*}=0$, $\alpha _{11}+\alpha _{12}+\alpha _{13}=1$
($\alpha _{11}-\frac{w_{1}}{w_{2}}\alpha _{13}-\alpha _{21}^{*}+\frac{w_{1}}{%
w_{2}}\alpha _{23}^{*}=1$) 
\begin{eqnarray*}
a_{x} &=&\left( C_{1}\right) ^{x-1}\left( C_{1}a_{0}+C_{2}\left(
b_{0}-a_{0}\right) x\right) +\alpha _{1}x+C_{2}\left( \alpha _{2}^{*}-\alpha
_{1}\right) \frac{x\left( x-1\right) }{2} \\
b_{x} &=&\left( C_{1}\right) ^{x-1}\left( C_{1}b_{0}+C_{5}\left(
a_{0}-b_{0}\right) x\right) +\alpha _{2}^{*}x+C_{5}\left( \alpha _{1}-\alpha
_{2}^{*}\right) \frac{x\left( x-1\right) }{2} \\
c_{x} &=&-\frac{w_{1}}{w_{2}}a_{x}+\frac{1}{w_{2}}b_{x}\text{.}
\end{eqnarray*}
\end{theorem}

In the special case, when (\ref{recf13}) is true, Theorem \ref{th_3rec_gen}
is reduced to Theorem \ref{th_3rec}.

\section{Conclusions and Open Problems}

We have considered the problem of finding a solution of simultaneous
recurrences. Section \ref{sec_gen_sim_rec} presents the general method of
arriving at a solution. Section \ref{sec_2_3_sim_rec} describes a special
method for solving matrix recurrences of order two and order three. This
method works only under the special conditions (specifically, when the sums
of coefficients of items in lines are equal) but seems to be more efficient.
It might be noted that perhaps, this method can be generalized to matrix
recurrences of order $n$.

It might also be of interest to investigate simultaneous recurrences with
equal sums of coefficients of items in columns.

\begin{lemma}
\label{lem_2rec_col}If 
\begin{equation*}
\left\{ 
\begin{array}{l}
a_{x}=\alpha _{11}a_{x-1}+\alpha _{12}b_{x-1}+\alpha _{1} \\ 
b_{x}=\alpha _{21}a_{x-1}+\alpha _{22}b_{x-1}+\alpha _{2}%
\end{array}%
\right.
\end{equation*}%
and 
\begin{equation*}
\alpha _{11}+\alpha _{21}=\alpha _{12}+\alpha _{22}\text{ ,}
\end{equation*}%
then

when $\alpha _{12}\neq -\alpha _{21}$, $\alpha _{11}-\alpha _{12}\neq 1$, $%
\alpha _{11}+\alpha _{21}\neq 1$%
\begin{eqnarray*}
a_{x} &=&\left( \alpha _{11}-\alpha _{12}\right) ^{x}a_{0}+\alpha
_{12}\left( a_{0}+b_{0}\right) \frac{\left( \alpha _{11}+\alpha _{21}\right)
^{x}-\left( \alpha _{11}-\alpha _{12}\right) ^{x}}{\alpha _{12}+\alpha _{21}}%
+ \\
&&\alpha _{1}\frac{\left( \alpha _{11}-\alpha _{12}\right) ^{x}-1}{\alpha
_{11}-\alpha _{12}-1}+\frac{\alpha _{12}\left( \alpha _{1}+\alpha
_{2}\right) }{\alpha _{12}+\alpha _{21}}\times \\
&&\left( \frac{\left( \alpha _{11}+\alpha _{21}\right) ^{x}-\alpha
_{11}-\alpha _{21}}{\alpha _{11}+\alpha _{21}-1}-\frac{\left( \alpha
_{11}-\alpha _{12}\right) ^{x}-\alpha _{11}+\alpha _{12}}{\alpha
_{11}-\alpha _{12}-1}\right) \\
b_{x} &=&\left( \alpha _{11}-\alpha _{12}\right) ^{x}b_{0}+\alpha
_{21}\left( a_{0}+b_{0}\right) \frac{\left( \alpha _{11}+\alpha _{21}\right)
^{x}-\left( \alpha _{11}-\alpha _{12}\right) ^{x}}{\alpha _{12}+\alpha _{21}}%
+ \\
&&\alpha _{2}\frac{\left( \alpha _{11}-\alpha _{12}\right) ^{x}-1}{\alpha
_{11}-\alpha _{12}-1}+\frac{\alpha _{21}\left( \alpha _{1}+\alpha
_{2}\right) }{\alpha _{12}+\alpha _{21}}\times \\
&&\left( \frac{\left( \alpha _{11}+\alpha _{21}\right) ^{x}-\alpha
_{11}-\alpha _{21}}{\alpha _{11}+\alpha _{21}-1}-\frac{\left( \alpha
_{11}-\alpha _{12}\right) ^{x}-\alpha _{11}+\alpha _{12}}{\alpha
_{11}-\alpha _{12}-1}\right) \text{.}
\end{eqnarray*}
\end{lemma}

\proof%
The proof is like that of Lemma \ref{lem_2rec}. 
\endproof%
\medskip

We intend to generalize Lemma \ref{lem_2rec_col} to matrix recurrences of
order $n$.

\end{document}